# Cloud-based Electronic Health Records for Real-time, Region-specific Influenza Surveillance.

**Authors:** M. Santillana[1,2,3], A. T. Nguyen[3], T. Louie[4], A. Zink[5], J. Gray[5], I. Sung, J. S. Brownstein[1,2].

**Author Affiliations:**

[1] Computational Health Informatics Program, Boston Children's Hospital, Boston, MA
[2] Harvard Medical School, Boston, MA
[3] Harvard School of Engineering and Applied Sciences, Cambridge, MA
[4] Harvard School of Public Health, Boston, MA
[5] athenaResearch at athenahealth, Watertown, MA

## Abstract

Accurate real-time monitoring systems of influenza outbreaks help public health officials make informed decisions that may help save lives. We show that information extracted from cloud-based electronic health records databases, in combination with machine learning techniques and historical epidemiological information, have the potential to accurately and reliably provide near real-time regional predictions of flu outbreaks in the United States.

## Introduction

Influenza is a leading cause of death in the United States (US), where up to 50,000 are killed each year by influenza-like illnesses (ILI) [1]. Therefore, monitoring, early detection, and prediction of influenza outbreaks are crucial to public health. Disease detection and surveillance systems provide epidemiologic intelligence that allows health officials to deploy preventive measures and help clinic and hospital administrators make optimal staffing and stocking decisions [2].

The US Centers for Disease Control and Prevention (CDC) monitors ILI in the US by gathering information from physicians' reports about patients with ILI seeking medical attention [3]. CDC's ILI data provides useful estimates of influenza activity; however, its availability has a known time lag of one to two weeks. This time lag is far from optimal since public health decisions need to be made based on information that is two weeks old. Systems capable of providing real-time estimates of influenza activity are, thus, critical.

Many attempts have been made to design methods capable of providing real-time estimates of ILI activity in the US by leveraging Internet-based data sources that could potentially measure ILI in an indirect manner [4, 5, 6, 7, 8, 9, 10, 11]. Google Flu Trends (GFT), a digital disease detection system that used Internet searches to predict ILI in the US, became the most widely used of these non-traditional methods [12] in the past few years. In August of 2015, GFT was shut down, opening opportunities for novel and reliable methods to fill the gap. Many lessons have been learned in the field of digital disease detection from multiple updates to GFT, proposed not only by Google, but also by other researchers [13, 14, 15, 16, 17, 18]. The performance of some of these updated models has substantially improved, for example, by including historical flu activity information as input, and by dynamically recalibrating the models, in order to not only incorporate the most up-to-date clinical information but also to adapt to new behavior in the population (for example, how Internet users search for health-related terms)[16, 17, 18]. Finally, very accurate real-time ILI estimates can now be produced, at the national-level, by combining disparate data streams in the US as shown in [19].

National-level predictions are hard to translate into actionable information that enable local health officials to make better decisions during a surge in clinical visits, for example [20]. As a consequence, accurate predictions at finer spatial resolutions are desirable. Cities within the same geographical region tend to have positively correlated estimates of epidemiologic model parameters such as the basic reproduction number, $R_0$, and such estimates tend to differ from region to region in the US [21]. Additionally, knowledge of the influenza level in one region, when used in conjunction with a network model drawn from social network analysis, has been shown to improve the accuracy of GFT's ILI forecasts nationally as well as in multiple regions [22]. Unfortunately, the accuracies of existing GFT-like systems degrade substantially at the regional and local resolutions [23].

Diverse studies have shown *retrospectively* the high correlation between aggregated data obtained from electronic health records (EHR) and flu syndromic surveillance systems [24, 25, 26], thus, suggesting the feasibility of using EHR data for disease tracking for both local and regional spatial resolutions. In the past, EHR data have not been used for real-time surveillance, due to reporting lag times of 1 to 2 weeks. Near real-time access to EHR records data would address this issue.

Here, we demonstrate EHR data collected and distributed in near real-time by an electronic health records and cloud services company, athenahealth, combined with historical patterns of flu activity using a suitable machine learning algorithm, can accurately predict real-time influenza activity (as reported by the CDC), at the regional scale in the United States. Additionally, we show that the signal to noise ratio in this data source is high. EHR data provides us with an "early count" of ILI activity in much the same way as exit polls enable us to forecast election results. We build a machine-learning model that optimally exploits the data by building a system as timely as the GFT used to be, yet as stable and reliable as CDC validated data sources. Although our model is capable of forecasting influenza levels weeks into the future, we decided to present here the real-time monitoring of ILI. We name our model ARES, which stands for AutoRegressive Electronic health record Support vector machine.

**Methods**

*Data*

Athenahealth is a provider of cloud-based services and mobile applications for medical groups and health systems (http://www.athenahealth.com). Its electronic health records, medical billing, and care coordination services are organized around a single cloud network, allowing for the collection of unique insights related to patient-provider encounter data for more than 72,000 healthcare providers in medical practices and health systems nationwide. This database includes claims data for over 64 million lives and electronic health records for over 23 million lives.

In collaboration with athenahealth, we obtained weekly total visit counts, flu vaccine visit counts, flu visit counts, ILI visit counts, and unspecified viral or ILI visit counts. The athenahealth ILI rates are based on visits to primary care providers on the athenahealth network, for the period between 6/28/09 and 10/15/15.

These providers see patients mostly in office-based settings (73.4% of the visits), though they also see patients in the following settings: inpatient hospital (11.3%), outpatient hospital (6.6%), nursing facility (1.5%), emergency room (1.4%), and other (5.8%). The age statistics for the athenahealth network are as follows: younger than 15 years old (15%), 15 to 24 years (6%), 25 to 44 years (14%), 45 to 64 years (29%), and 65 years or older (36%). The place of service and age distribution statistics are close to the statistics reported by the CDC's National Ambulatory Medical Care Survey (NAMCS). This suggests that medical visits to athenahealth's medical provider network provide a good representation of ambulatory care, including flu activity, in the US.

There are two main differences to note between the athenahealth and NAMCS statistics. First, providers in the athenahealth network see patients that are slightly older than average, though re-weighting the data change the athenahealth ILI rates very little. Second, emergency departments are underrepresented in the athenahealth network, causing the raw athenahealth %ILI estimates to be consistently lower than the values reported by the CDC. This indicates that the athenahealth data could be adapted to build specialized models that mimic ILI rates.

To develop these models, we used the following list of athenahealth variables as input: *Flu Vaccine Visit Count*, which accounts for the number of visits where a flu vaccine was administered; *Flu Visit Count*, the number of visits where the patient had a flu diagnosis; *ILI Visit Count*, the number of visits where the patient had either a flu diagnosis or a fever diagnosis with an accompanying sore throat or cough diagnosis; and *Unspecified Viral or ILI Visit Count*, the number of visits where the patient had either an unspecified viral diagnosis, a flu diagnosis, or a fever diagnosis with an accompanying sore throat or cough diagnosis. It is important to note that *Unspecified Viral or ILI Visit Count* contains *ILI Visit Count,* which itself contains *Flu Visit Count.* In [19] only *ILI Visit Count* was used as an independent variable.

We also obtained the national and regional ILI weekly values from the CDC (gis.cdc.gov/grasp/fluview/fluportaldashboard.html), for the same time period, to use as both a comparator as well as to provide historical independent variables for our models. We note that using weekly information from reports published by the CDC as a gold standard for US national and regional influenza activity may have limitations. An evaluation of the CDC's data collection approach has been published and improvements suggested in [27].

We obtained historical GFT data for the same time period to use as a comparison from the Google Flu Trends website (http://www.google.org/flutrends).

All of the data used was downloaded on Oct 15, 2015 and all of the experiments were conducted in Python 2.7.

*Analysis*

We built a collection of regional models that dynamically re-calibrate every week in order to optimally include all of the available data up to the week of prediction. This dynamic recalibration is inspired by data assimilation techniques used in industrial and financial time series forecasting [28] and weather forecasting [29]. The independent variables used by ARES to produce real-time estimates of ILI activity at time $t$ include: *Unspecified Viral or ILI Visit Count, ILI Visit Count,* and *Flu Visit Count* from athenahealth, for weeks *(t-2), (t-1) and* (*t).* We also incorporated (autoregressive) historical information from the unweighted percentage ILI estimates obtained from the CDC for weeks *(t-2)* and *(t-1)*.

Our ARES models map the variables described above into a %ILI real-time estimate, using a support vector machine (SVM) model [30]. Support vector machine models are similar to multivariate regression models with the important difference that non-linear functions can be learned via the kernel trick, which implicitly maps the independent variables to a higher dimensional feature space. The independent variables can even be mapped to an infinite dimensional feature space with the use of a radial basis function kernel. SVM models are fitted by minimizing an epsilon insensitive cost function where errors of magnitude less than epsilon are ignored by the cost function, leading to better generalization of the learned model. The SVM kernel type, margin width, and regularization hyper parameters were chosen via cross-validation on the training data.

For comparison purposes, we produced real-time estimates using two baseline methodologies: (a) a dynamically-trained autoregressive model that only used historical CDC information, called AR(2) throughout the paper; and (b) a dynamically-trained linear model that used athenahealth's %ILI information onto CDC's ILI, as introduced in [19]. The independent variables used to produce ILI estimates for the week at time t are: (a) for the AR(2) model: CDC's ILI for weeks *(t-2)* and *(t-1),* and (b) for the dynamically trained linear model: the value of athenahealth's ILI value at time *t*.

**Results**

The training period for our first prediction consisted of data from 6/28/2009 through 1/1/2012 for all the implemented models. Thus, our first real-time estimate of ILI was produced for the week of 1/8/2012. Time series of real-time out-of-sample estimates using ARES, the AR(2) model, and the linear model, were generated up to and including the week of June 28, 2015. Figure 1 shows the national level real-time estimates produced by ARES and the target CDC ILI signal. GFT estimates, as well as the estimates produced by the two baseline models described in the previous section, are included for comparison purposes. Figure 2 shows the same results but at a regional resolution for each of the 10 regions defined by the Health and Human Services (HHS).

Table 1 shows the accuracy metrics, as defined in [19], between the models' predictions and the target signal, CDC's ILI. ARES provides accurate real-time estimates of ILI activity in all ten HHS regions as well as at the national level. The average Pearson correlation across all ten regions is 0.972, and the national Pearson correlation is 0.996. The average root mean square error (RMSE) across all ten regions is 0.261, and the national RMSE is 0.10. The average relative RMSE across all ten regions is 18.28%, and the national relative RMSE is 4.63%. The majority of the regional error is due to Region 7 estimates.

In order to understand the predictive power of each of the independent variables to estimate CDC's ILI, we plotted the values of the coefficient associated with each variable in the (dynamically-trained) linear models as a function of time. These values are presented in the multiple heatmaps of Figure 4. These heatmaps show that the variables with highest predictive power are CDC's ILI value during the previous week of prediction (t-1), athenahealth's viral visit counts during the week of prediction (t), and athenahealth's ILI during the week of prediction (t).

**Discussion**

In this study we have shown that EHR data in combination with historical patterns of flu activity and a robust dynamical machine-learning algorithm, are capable of accurately predicting real-time influenza activity at the national and regional scales in the US. Table 1 shows that ARES is capable of predicting national ILI activity with an almost ten-fold reduction in the average error (RMSE) reported by the now discontinued Google Flu Trends system, during our study period. In most regions, 5-10 fold (RMSE) error reductions are observed when comparing to GFT historical regional ILI estimates. Substantial improvements in Pearson correlation are observed from historical GFT predictions and ARES. Nationally, for example, GFT's Pearson correlation is 0.91 while ARES's is 0.996. The most significant improvement on Pearson correlation (from GFT's 0.734 to ARES's 0.99) happens in region 4, and the worst (from GFT's 0.873 to ARES's 0.938) occurs in region 7. Overall, ARES consistently outperforms GFT's historical estimates in all statistics.

While the ability of athenahealth's ILI data to predict CDC's ILI *nationally* was established using a dynamically-trained linear model in [19], here we show that incorporating CDC's ILI historical information and more EHR information, using a suitable machine learning methodology, can improve predictions substantially. Specifically, as shown in the heatmaps of Figure 4, adding CDC's ILI for the previous week of prediction, and the variable associated to athenahealth's *viral visits* during the week of prediction improve results.

Predictions using ARES lead to 2-3 fold error reductions in the national and regional predictions, when compared to the dynamic linear model introduced in [19] for mapping athenahealth's ILI data onto CDC's ILI. Pearson correlations improved nationally and across regions when comparing ARES to the dynamic linear model in [19]. The most substantial improvement (from 0.756 to 0.938) happened in region 7, and the mildest (from 0.967 to 0.971) in region 1.

Models that use only historical information to predict future ILI typically show high Pearson correlation and not very large (RMSE) error values when compared with CDC's ILI; however, they consistently show lags of 1 to 2 weeks with respect to the observed CDC's ILI values, making them systematically inaccurate. This happens to our baseline AR(2) model implementation across regions, as shown in Figures 1 and 2. Interestingly, when using ARES to combine CDC's ILI historical information with EHR data, predictions improved. The improvement of predictions when historical information is added has been previously observed in methodologies that use Google searches [17] and Twitter [31] to estimate flu activity. Intuitively speaking, historical information maintains the estimates within a reasonable range, while EHR data improves the responsiveness of the model to changes. This responsiveness and stability can be seen when comparing the estimates of ARES with the two baseline methodologies in Figures 1 and 2.

Specifically, using ARES improved Pearson correlation (from 0.958 to 0.996) and reduced the error (RMSE) three-fold for the national level when compared to the autoregressive, AR(2), model. Pearson correlation across regions was generally improved, with the largest improvement in region 3 (from 0.932 to 0.987) and the mildest taking place in region 2 (from 0.96 to 0.978). The average error (RMSE) generally improved, with the greatest performance in region 4, where more than a two-fold reduction was achieved, and the mildest reduction in region 9, where a 20% reduction in error was achieved.

The only region where the combination of historical CDC data and EHR data did not lead to improvements when compared to the AR(2) model was region 7, where correlations went from 0.958 to 0.938, and the average error (RMSE) went from 0.4 to 0.51 (%ILI). This may be explained by the fact that athenahealth manages very few facilities in the Midwestern United States.

We would like to note that in this study we used the *revised* version of historical CDC ILI reports to dynamically train all of our models. While this *revised* information was not technically available at the time of predictions (it was probably available a week or two later), our main point in this work is to show the methodological advantages of combining historical data with EHR data to improve predictions. This goal is achieved since all models (including the baseline ones) were trained with the same *revised* data. However, our comparison with the historical GFT values may not be strictly fair. It is important to highlight that this decision was not taken lightly. Our experience training flu prediction models [17,19] has shown us that the results of our predictions training with only the historically available (revised and unrevised) CDC's ILI values, at the time of prediction, versus training with only revised CDC's ILI values change minimally. Specifically, as we have reported in [17], the performance of our models is almost unchanged (correlation slightly improving from 0.985 to 0.986, for example) for national predictions when using only *revised* historical CDC reports to train them, as opposed to using only the available ones (revised and unrevised) at the time. See the differences between Table 1 and Table S1 in [17], for example.

The discrepancies between the predictions using ARES and the observed CDC values, as captured by Pearson correlation and RMSE values, are comparable to those shown in [19] where an ensemble approach was used to combine multiple models based on disparate data sources to obtain optimal predictions. The resulting ensemble predictions outperformed any single-source flu prediction model. We suspect that using ARES as input in a new ensemble approach will lead to improved predictions.

Finally, ARES avoids many of the limitations faced by most non-traditional digital disease surveillance systems, such as a lack of specificity, while providing the main advantage of non-traditional systems, timeliness. The most limiting aspect of ARES is that it relies on electronic health records, data that are rarely publically available. At the time of the writing of this report, athenahealth data is provided to several groups of flu researchers around the country but is not publically available. Future work includes integrating ARES into a network model and testing the accuracy of ARES at the state and city levels, in other countries, as well as on other communicable diseases.

**Conclusion**

We have shown that EHR data in combination with historical patterns of flu activity and a robust dynamical machine-learning algorithm provide a novel and promising way of monitoring infectious diseases at the national and local level. Our methodology provides timely predictions with the accuracy and specificity of sentinel systems like the CDC's ILI surveillance network. This demonstrates the value of cloud-based electronic health records databases for public health at the local level.

**Author contributions**

MS, JSB, IS, and JG conceived the research. IS, AZ and JG collected and prepared the aggregated electronic health records data. MS, ATN, and TL performed the analysis and developed the predictive algorithms. MS and ATN wrote the main manuscript text. TL prepared the figures and tables. All authors reviewed the manuscript.

**Competing financial interests**
IS, AZ, and JG were employed by athenahealth at the time this research was performed. MS, ATN, TL, and JSB declare no potential conflict of interest.

## Figures

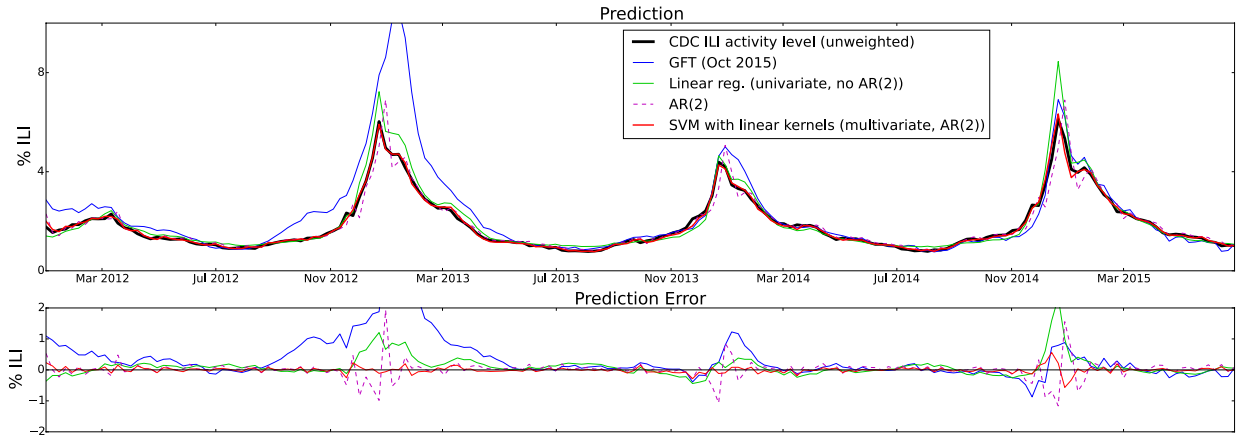

**Figure 1**: The CDC's ILI estimates, GFT estimates, baseline linear regression and AR(2) autoregressive model estimates, and ARES estimates are displayed as a function of time for the national level on the top panel. The errors associated with GFT, the linear regression and autoregressive model baselines, and ARES are shown on the bottom panel.

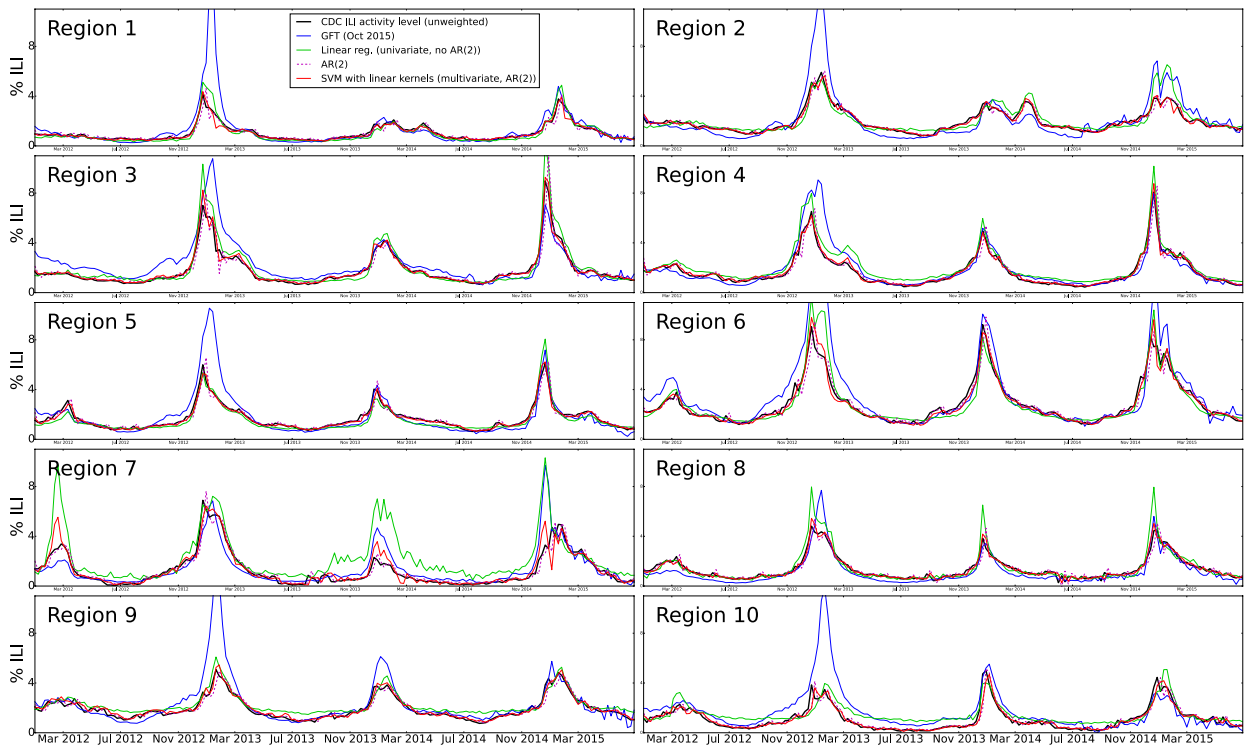

**Figure 2**: The CDC's ILI estimates, GFT estimates, baseline linear regression and AR(2) autoregressive model estimates, and ARES estimates are displayed as a function of time for each of the 10 US regions defined by the HHS.

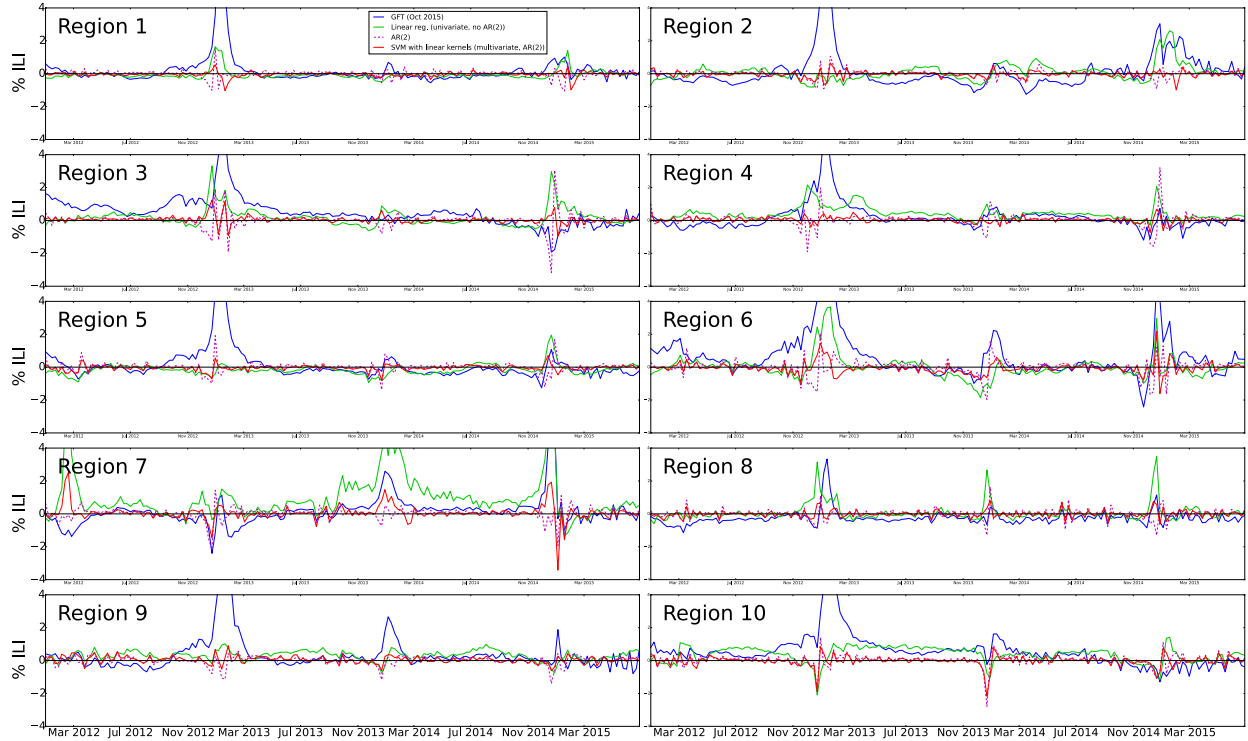

**Figure 3**: The errors associated with GFT, the linear regression and AR(2) autoregressive model baselines, and ARES are displayed as a function of time for each of the 10 US regions defined by the HHS.

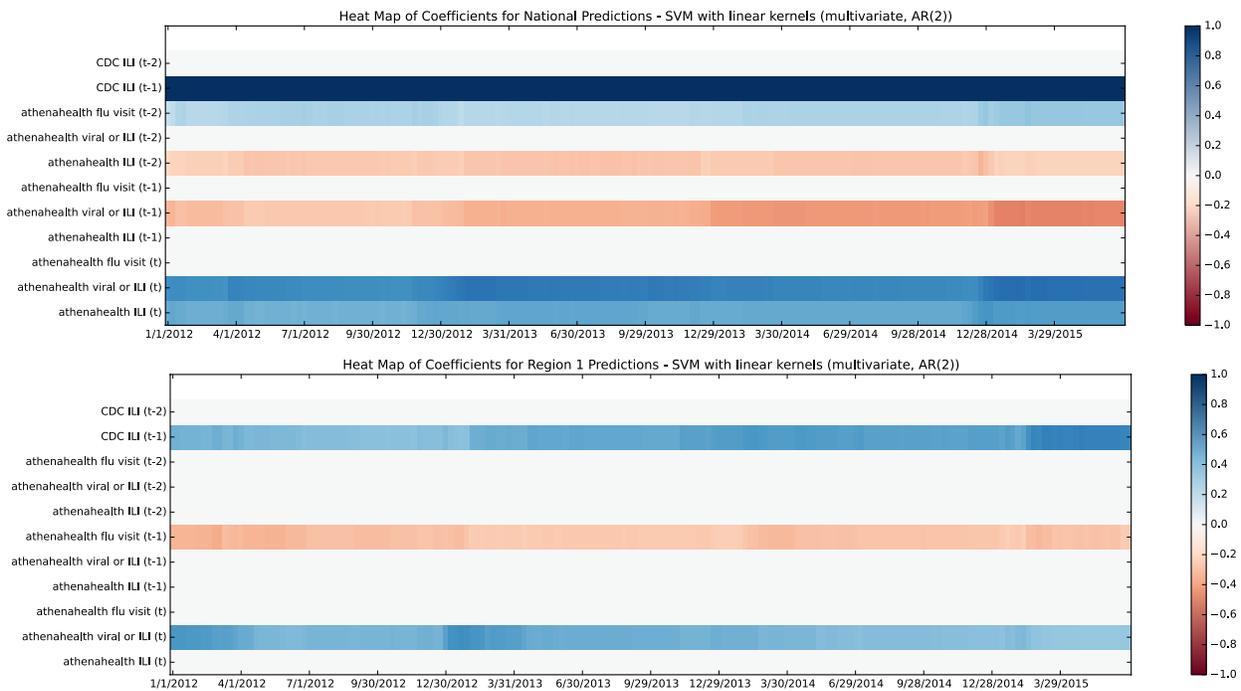

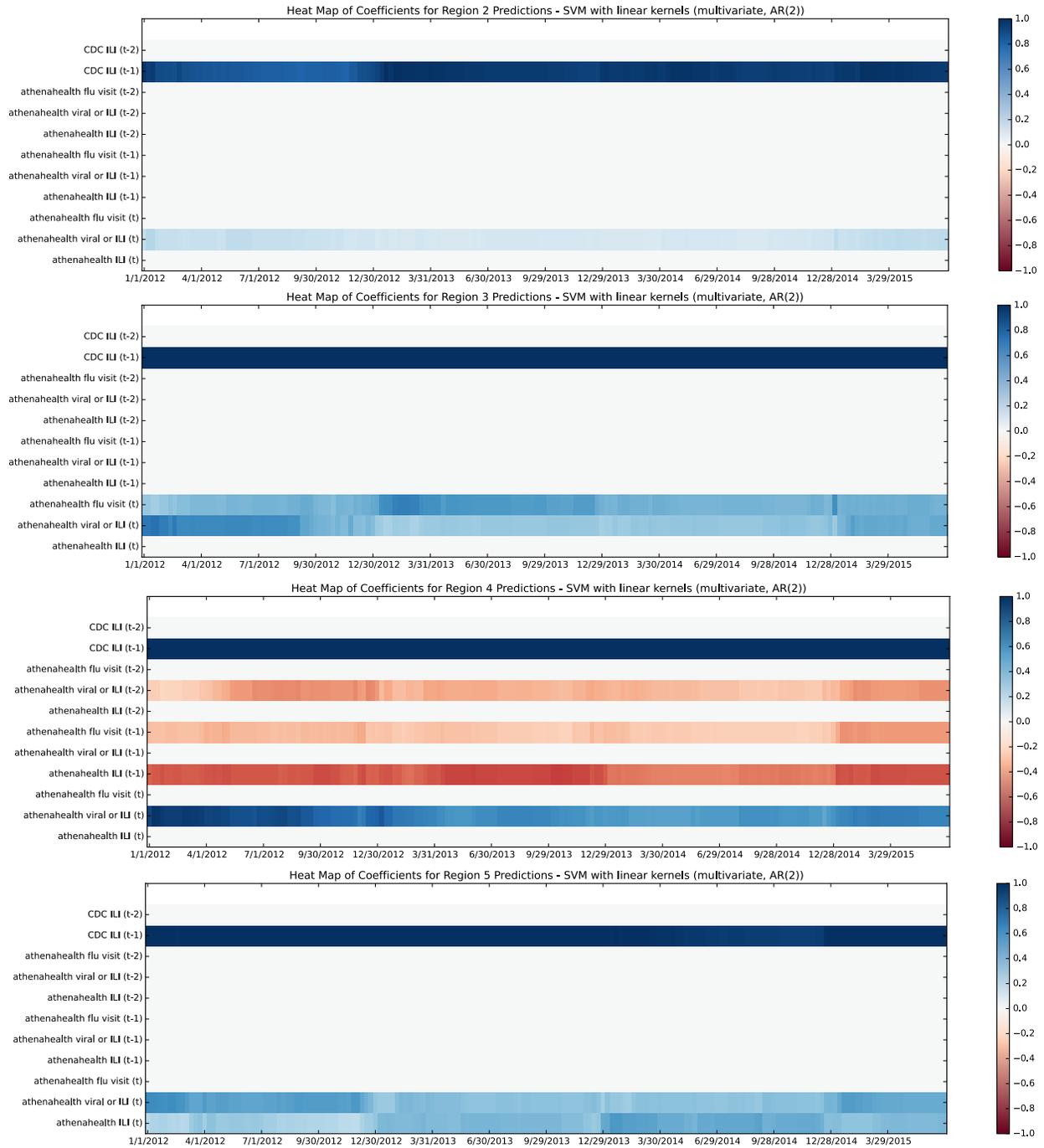

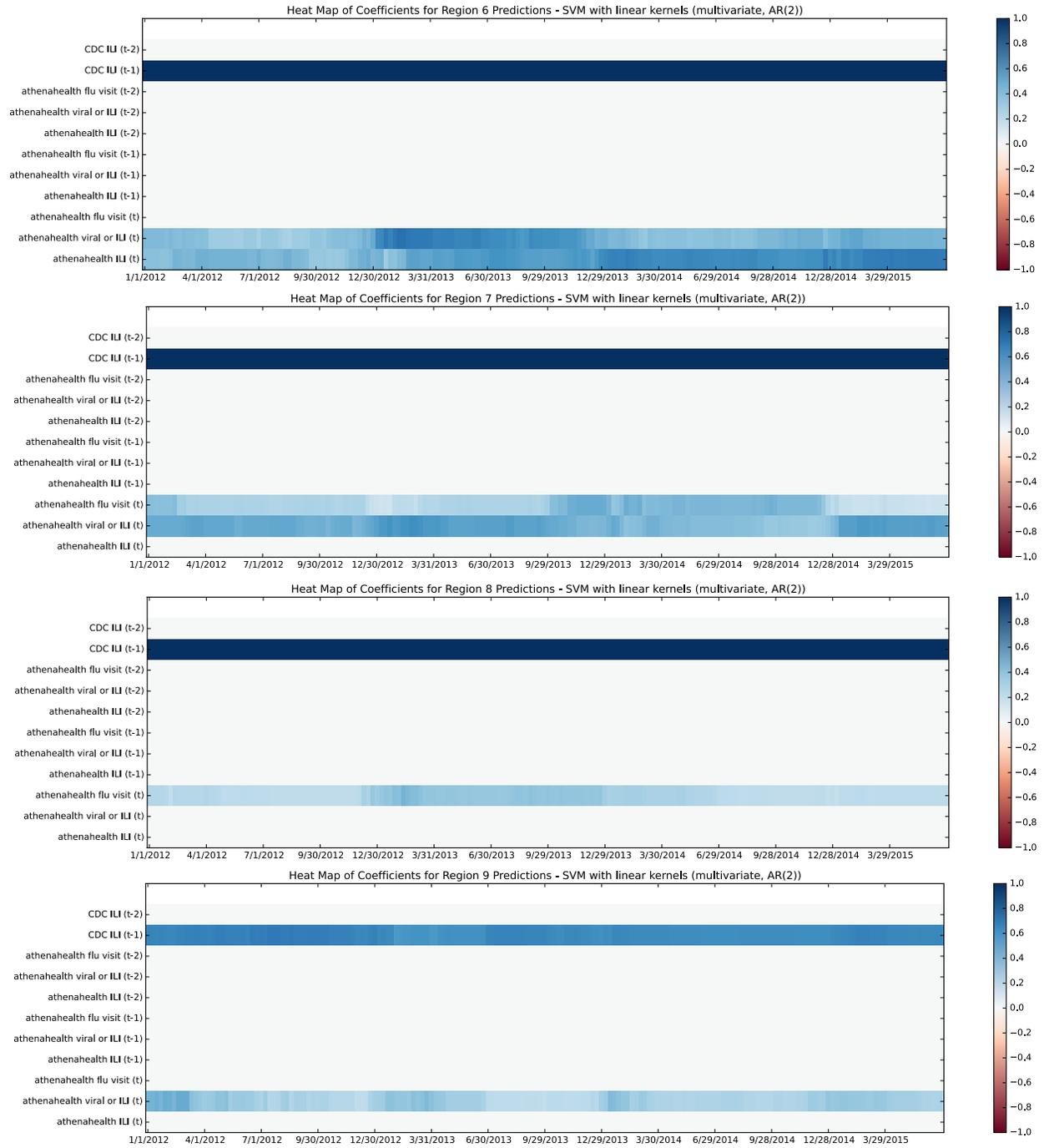

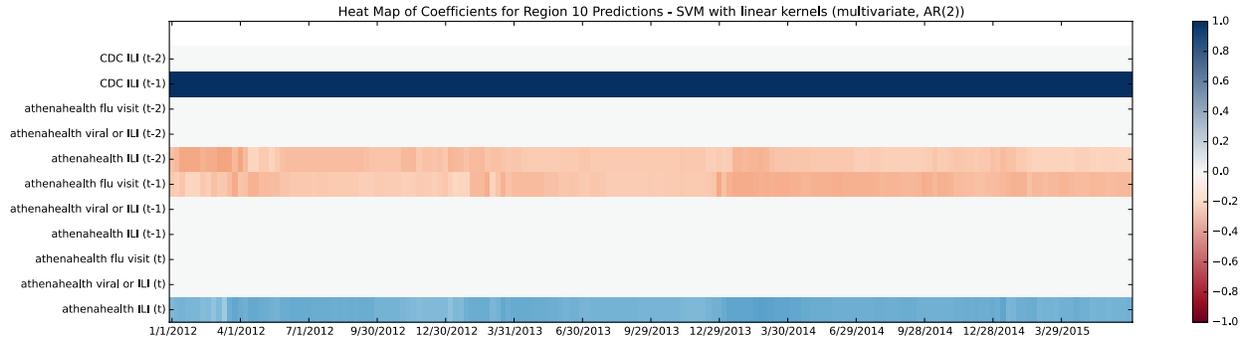

**Figure 4.** Values of the linear coefficients associated with each independent variable as a function of time for the national and the 10 regional prediction models.

| Algorithm | RMSE | Rel. RMSE (%) | Correlation |
|---|---|---|---|
| **National** | | | |
| GFT | 0.96 | 30.82% | 0.910 |
| Linear (univariate) | 0.33 | 11.65% | 0.984 |
| AR(2) | 0.31 | 10.05% | 0.958 |
| SVM (linear) + AR(2) | 0.10 | 4.63% | 0.996 |
| **Region 1** | | | |
| GFT | 1.74 | 137.19% | 0.796 |
| Linear (univariate) | 0.29 | 21.05% | 0.967 |
| AR(2) | 0.25 | 18.84% | 0.928 |
| SVM (linear) + AR(2) | 0.17 | 11.92% | 0.971 |
| **Region 2** | | | |
| GFT | 1.28 | 55.93% | 0.824 |
| Linear (univariate) | 0.52 | 21.72% | 0.887 |
| AR(2) | 0.26 | 10.36% | 0.960 |
| SVM (linear) + AR(2) | 0.20 | 8.48% | 0.978 |
| **Region 3** | | | |
| GFT | 1.10 | 40.10% | 0.845 |
| Linear (univariate) | 0.56 | 21.34% | 0.982 |
| AR(2) | 0.50 | 15.51% | 0.932 |
| SVM (linear) + AR(2) | 0.23 | 9.57% | 0.987 |
| **Region 4** | | | |
| GFT | 1.26 | 75.89% | 0.734 |
| Linear (univariate) | 0.56 | 41.10% | 0.971 |
| AR(2) | 0.44 | 17.00% | 0.939 |
| SVM (linear) + AR(2) | 0.18 | 10.83% | 0.990 |
| **Region 5** | | | |
| GFT | 0.95 | 35.72% | 0.894 |
| Linear (univariate) | 0.36 | 18.44% | 0.951 |
| AR(2) | 0.33 | 14.16% | 0.942 |
| SVM (linear) + AR(2) | 0.18 | 9.12% | 0.984 |
| **Region 6** | | | |
| GFT | 1.58 | 48.14% | 0.800 |
| Linear (univariate) | 0.71 | 16.75% | 0.939 |
| AR(2) | 0.51 | 12.74% | 0.958 |
| SVM (linear) + AR(2) | 0.37 | 9.56% | 0.980 |
| **Region 7** | | | |
| GFT | 3.02 | 556.64% | 0.873 |
| Linear (univariate) | 1.60 | 297.65% | 0.756 |
| AR(2) | 0.40 | 62.29% | 0.958 |
| SVM (linear) + AR(2) | 0.51 | 54.13% | 0.938 |
| **Region 8** | | | |
| GFT | 0.75 | 37.87% | 0.929 |
| Linear (univariate) | 0.54 | 28.01% | 0.942 |
| AR(2) | 0.35 | 37.70% | 0.930 |
| SVM (linear) + AR(2) | 0.22 | 27.60% | 0.974 |
| **Region 9** | | | |
| GFT | 1.06 | 55.68% | 0.812 |
| Linear (univariate) | 0.43 | 33.27% | 0.944 |
| AR(2) | 0.28 | 12.20% | 0.951 |
| SVM (linear) + AR(2) | 0.22 | 11.97% | 0.973 |
| **Region 10** | | | |
| GFT | 1.82 | 234.57% | 0.732 |
| Linear (univariate) | 0.60 | 170.80% | 0.867 |
| AR(2) | 0.39 | 38.94% | 0.916 |
| SVM (linear) + AR(2) | 0.33 | 29.64% | 0.941 |

**Table 1.** Similarity metrics between ARES and CDC's ILI for all geographic regions. For comparison purposes, we have included GFT's historical predictions, and the two baseline models: dynamic linear regression (mapping athenahealth's ILI onto CDC's ILI), and a two term autoregressive model, AR(2).

## Appendix 1

The US Department of Health and Human Services (HHS) divides the US into the following 10 regions:

Region 1: *Connecticut, Maine, Massachusetts, New Hampshire, Rhode Island, and Vermont*
Region 2: *New Jersey, New York, Puerto Rico, and the U.S. Virgin Islands*
Region 3: *Delaware, District of Columbia, Maryland, Pennsylvania, Virginia, and West Virginia*
Region 4: *Alabama, Florida, Georgia, Kentucky, Mississippi, North Carolina, South Carolina, and Tennessee*
Region 5: *Illinois, Indiana, Michigan, Minnesota, Ohio, and Wisconsin*
Region 6: *Arkansas, Louisiana, New Mexico, Oklahoma, and Texas*
Region 7: *Iowa, Kansas, Missouri, and Nebraska*
Region 8: *Colorado, Montana, North Dakota, South Dakota, Utah, and Wyoming*
Region 9: *Arizona, California, Hawaii, and Nevada*
Region 10: *Alaska, Idaho, Oregon, and Washington*